\documentstyle[12pt]{article}

\def\noi{\noindent}

\def\nqq{\hspace{-2em}}

\def\beq#1{\begin{equation}\label{#1}}
\def\eeq{\end{equation}}
\def\ber#1{\begin{eqnarray}\label{#1} \nqq}
\def\eer{\end{eqnarray}}
\def\nn{\nonumber}
\newcommand{\bear}[1]{\begin{eqnarray}\label{#1}}
\newcommand{\ear}{\end{eqnarray}}

\catcode`\@=11 \@addtoreset{equation}{section}\catcode`\@=12

\newcommand{\R}{\mbox{\bf R}}

\newcommand{\N}{\mbox{\bf N}}

\newcommand{\sign}{\mathop{\rm sign}\nolimits}

\newcommand{\eps}{\varepsilon}
\newcommand{\tri}{\triangle}

\newcommand{\p}{\partial}

\newcommand{\fnm}{\footnotemark}
\newcommand{\fnt}{\footnotetext}

\begin{document}

\begin{center}
\large\bf
P-BRANE BLACK HOLES  FOR GENERAL INTERSECTIONS
\\[15pt]
\normalsize\bf V.D. Ivashchuk\fnm[1]\fnt[1]{ivas@rgs.phys.msu.su},
and
V.N. Melnikov
\fnm[2]\fnt[2]{melnikov@rgs.phys.msu.su} \\[10pt]

\it Center for Gravitation and Fundamental Metrology,
VNIIMS, 3/1 M. Ulyanovoy Str.,
Moscow 117313, Russia  and\\
Institute of Gravitation and Cosmology, PFUR,
Michlukho-Maklaya Str. 6, \\ Moscow 117198, Russia

\end{center}

\vspace{15pt}

\small\noi

\begin{abstract}

Black hole  generalized $p$-brane  solutions  for a wide class of
intersection rules are presented. The solutions are defined on a
manifold that contains a product of $n - 1$ Ricci-flat internal
spaces. They are defined up to moduli functions $H_s = H_s(R)$
obeying a non-linear differential equations  (equivalent to
Toda-type equations) with  certain boundary conditions imposed.
Using conjecture on polynomial structure of $H_s$  for
intersections related to Lie algebras, new $A_2$-dyon solutions
are obtained. Two examples of these $A_2$-dyon solutions, i.e.
dyon in  $D = 11$ supergravity with $M2$ and $M5$ branes
intersecting at a point and dyon in Kaluza-Klein theory, are
considered.

\vspace{1cm}


\vspace{1mm}


\end{abstract}

\vspace{10cm}

\pagebreak

\normalsize

\section{Introduction}

At present there exists some interest to  $M$-theory
(see, for example, \cite{M-th1}-\cite{M-th2}).
This theory is ``supermembrane'' analogue of
superstring models \cite{GSW} in $D=11$. The low-energy limit of
$M$-theory after a dimensional reduction leads to models governed by a
Lagrangian containing a metric, fields of  forms and scalar fields.
These models contain a large variety of the so-called
$p$-brane solutions (see \cite{St}-\cite{IK} and references
therein).

In \cite{IMC} it was shown that after
the dimensional reduction on the
manifold $M_0\times M_1\times\dots\times M_n$  when the composite
$p$-brane ansatz for fields of forms
is considered the problem is reduced to the gravitating
self-interacting $\sigma$-model with certain constraints imposed. (For
electric $p$-branes see also \cite{IM0,IM,IMR}.) This representation
may be considered as a tool for obtaining different solutions
with intersecting $p$-branes. In \cite{IMC,IMR,IMBl,GrI}
the Majumdar-Papapetrou type solutions  were obtained (for non-composite
case see \cite{IM0,IM}). These solutions correspond to Ricci-flat
factor-spaces $(M_i,g^i)$, ($g^i$ is metric on $M_i$)
$i=1,\dots,n$.They were also
generalized to the case of Einstein internal spaces \cite{IMC}. (Earlier
some special classes of these solutions were considered in
\cite{Ts1,PT,GKT,AR,AEH,AIR}.) The solutions take place, when
certain (block-)orthogonality relations (on couplings parameters,
dimensions of "branes", total dimension) are imposed. In this situation a
class of cosmological and spherically-symmetric solutions was obtained
\cite{IMJ,Br2,IMJ2}. The solutions with a horizon were studied
in details in \cite{CT,AIV,Oh,IMJ,BIM,IMBl,Br2,CIM}.

Here we present a family of $p$-brane black hole  solutions
with (next to) arbitrary intersections (see Sect. 2).
These black hole solutions are
governed by moduli functions $H_s = H_s(R)$ obeying a set of second order
non-linear differential equations  with some boundary relations
imposed. Some general features
of these black holes (e.g. ``single-time'' and ``no-hair'' theorems)
were predicted earlier in \cite{Br2}.
We suggested a conjecture: the moduli functions
$H_s$ are polynomials when intersection rules correspond to semisimple Lie
algebras. The conjecture was
confirmed by special  black-hole ``block -orthogonal'' solutions considered
earlier in \cite{Br,IMBl,CIM,IMJ2}.  An analogue of this conjecture for
extremal black holes was considered earlier in \cite{LMMP}.

In Sect. 3 explicit formulas for the solution
corresponding to the Lie algebra $A_2$ are obtained.  These formulas are
illustrated by two examples of $A_2$-dyon solutions: a dyon in $D = 11$
supergravity (with $M2$ and $M5$ branes intersecting at a point)
and Kaluza-Klein dyon.


\section{$p$-brane black hole solutions}

We consider a  model governed by
the action \cite{IMC}
\ber{1.1}
S=\int d^Dx \sqrt{|g|}\biggl\{R[g]-h_{\alpha\beta}g^{MN}\p_M\varphi^\alpha
\p_N\varphi^\beta-\sum_{a\in\tri}\frac{\theta_a}{n_a!}
\exp[2\lambda_a(\varphi)](F^a)^2\biggr\}
\eer
where $g=g_{MN}(x)dx^M\otimes dx^N$ is a metric,
$\varphi=(\varphi^\alpha)\in\R^l$ is a vector of scalar fields,
$(h_{\alpha\beta})$ is a  constant symmetric
non-degenerate $l\times l$ matrix $(l\in \N)$,
$\theta_a=\pm1$,
\beq{1.2a}
F^a =    dA^a
=  \frac{1}{n_a!} F^a_{M_1 \ldots M_{n_a}}
dz^{M_1} \wedge \ldots \wedge dz^{M_{n_a}}
\eeq
is a $n_a$-form ($n_a\ge1$), $\lambda_a$ is a
1-form on $\R^l$: $\lambda_a(\varphi)=\lambda_{\alpha a}\varphi^\alpha$,
$a\in\tri$, $\alpha=1,\dots,l$.
In (\ref{1.1})
we denote $|g| =   |\det (g_{MN})|$,
\beq{1.3a}
(F^a)^2_g  =
F^a_{M_1 \ldots M_{n_a}} F^a_{N_1 \ldots N_{n_a}}
g^{M_1 N_1} \ldots g^{M_{n_a} N_{n_a}},
\eeq
$a \in \tri$. Here $\tri$ is some finite set.
In the models with one time all $\theta_a =  1$
when the signature of the metric is $(-1,+1, \ldots, +1)$.

We obtained a new family of (black hole) solutions
to field equations corresponding to the action
(\ref{1.1}) (for derivation of these solutions see \cite{IMc}).
These solutions are  defined on the manifold
\beq{1.2}
M =    (R_{0}, + \infty)
\times (M_1 = S^{d_1}) \times (M_2 = \R) \times  \ldots \times M_n,
\eeq
and have the following form
\bear{2.30}
g= \Bigl(\prod_{s \in S} H_s^{2 h_s d(I_s)/(D-2)} \Bigr)
\biggl\{ f^{-1} dR \otimes dR
+ R^2  d \Omega^2_{d_1}  \\ \nn
-  \Bigl(\prod_{s \in S} H_s^{-2 h_s} \Bigr)
f  dt \otimes dt
+ \sum_{i = 3}^{n} \Bigl(\prod_{s\in S}
  H_s^{-2 h_s \delta_{iI_s}} \Bigr) g^i  \biggr\},
\\  \label{2.31}
\exp(\varphi^\alpha)=
\prod_{s\in S} H_s^{h_s \chi_s \lambda_{a_s}^\alpha},
\\  \label{2.32a}
F^a= \sum_{s \in S} \delta^a_{a_s} {\cal F}^{s},
\ear
where $f =1 - 2\mu/R^{\bar d}$,
\beq{2.32}
{\cal F}^s= - \frac{Q_s}{R^{d_1}}
\left( \prod_{s' \in S}  H_{s'}^{- A_{s s'}} \right) dR \wedge\tau(I_s),
\eeq
$s\in S_e$,
\beq{2.33}
{\cal F}^s= Q_s \tau(\bar I_s),
\eeq
$s\in S_m$.
Here $Q_s \neq 0$ ($s\in S$) are charges, $R_0 >0$,
$R_0^{\bar d} =2 \mu > 0$, $\bar d = d_1 -1$.
In  (\ref{2.30})
$g^i=g_{m_i n_i}^i(y_i) dy_i^{m_i}\otimes dy_i^{n_i}$
is a Ricci-flat  metric on $M_{i}$, $i=  3,\ldots,n$
and
\beq{1.11}
\delta_{iI}=  \sum_{j\in I} \delta_{ij}
\eeq
is the indicator of $i$ belonging
to $I$: $\delta_{iI}=  1$ for $i\in I$ and $\delta_{iI}=  0$ otherwise.
Let
$g^2 = -dt \otimes dt$,
and $g^1 = d \Omega_{d_1}$ be a canonical metric
on unit sphere $M_1 =S^{d_1}$,

The  $p$-brane  set  $S$ is by definition
\ber{1.6}
S=  S_e \cup S_m, \quad
S_v=  \cup_{a\in\tri}\{a\}\times\{v\}\times\Omega_{a,v},
\eer
$v=  e,m$ and $\Omega_{a,e}, \Omega_{a,m} \subset \Omega$,
where $\Omega =   \Omega(n)$  is the set of all non-empty
subsets of $\{ 2, \ldots,n \}$, i.e.
all $p$-branes do not ``live'' in  $M_1$.

Any $p$-brane index $s \in S$ has the form
$s =   (a_s,v_s, I_s)$,
where
$a_s \in \tri$, $v_s =  e,m$ and $I_s \in \Omega_{a_s,v_s}$.
The sets $S_e$ and $S_m$ define electric and magnetic $p$-branes
correspondingly. In (\ref{2.31}) $\chi_s  =   +1, -1$
for $s \in S_e, S_m$ respectively. All $p$-branes
contain the time manifold $M_2 = \R$, i.e.
\ber{1.7a}
2 \in I_s, \qquad \forall s \in S.
\eer

All the  manifolds $M_{i}$, $i > 2$, are assumed to be oriented and
connected and  the volume $d_i$-forms
\beq{1.12}
\tau_i  \equiv \sqrt{|g^i(y_i)|}
\ dy_i^{1} \wedge \ldots \wedge dy_i^{d_i},
\eeq
are well--defined for all $i=  1,\ldots,n$.
Here $d_{i} =   {\rm dim} M_{i}$, $i =   1, \ldots, n$,
$d_1 > 1$, $d_2 = 1$, and for any
 $I =   \{ i_1, \ldots, i_k \} \in \Omega$, $i_1 < \ldots < i_k$,
we denote
\ber{1.13}
\tau(I) \equiv \tau_{i_1}  \wedge \ldots \wedge \tau_{i_k},
\qquad
d(I)  =  \sum_{i \in I} d_i.
\eer
Forms ${\cal F}^s$  correspond to electric
and magnetic $p$-branes for $s\in S_e, S_m$ respectively.
In (\ref{2.33})
\beq{1.13a}
\bar I= \{1,\ldots,n\}\setminus I.
\eeq

The parameters  $h_s$ appearing in the solution
satisfy the relations
\beq{1.16}
h_s = K_s^{-1}, \qquad  K_s = B_{s s},
\eeq
where
\ber{1.17}
B_{ss'} =
d(I_s\cap I_{s'})+\frac{d(I_s)d(I_{s'})}{2-D}+
\chi_s\chi_{s'}\lambda_{\alpha a_s}\lambda_{\beta a_{s'}}
h^{\alpha\beta},
\eer
$s, s' \in S$, with $(h^{\alpha\beta})=(h_{\alpha\beta})^{-1}$
and $D =   1 + \sum_{i =   1}^{n} d_{i}$.
Here we assume that
\beq{1.17a}
({\bf i}) \qquad B_{ss} \neq 0,
\eeq
for all $s \in S$, and
\beq{1.18b}
({\bf ii}) \qquad {\rm det}(B_{s s'}) \neq 0,
\eeq
i.e. the matrix $(B_{ss'})$ is a non-degenerate one.

Let consider the matrix
\beq{1.18}
(A_{ss'}) = \left( 2 B_{s s'}/B_{s' s'} \right).
\eeq
Here  some ordering in $S$ is assumed.

Functions $H_s = H_s(z) > 0$, $z = 2\mu/R^{\bar d} \in (0,1)$
obey the equations
\beq{3.1}
 \frac{d}{dz} \left( \frac{(1-z)}{H_s} \frac{d H_s}{dz} \right) = B_s
\prod_{s' \in S}  H_{s'}^{- A_{s s'}},
\eeq
equipped with the boundary conditions
\bear{3.2a}
H_{s}(1 - 0) = H_{s0} \in (0, + \infty), \\
\label{3.2b}
H_{s}(+ 0) = 1,
\ear
$s \in S$. Here $B_{s} = K_s \eps_s Q_s^2/(2 \bar d \mu)^2$
and
\beq{1.22}
\eps_s=(-\eps[g])^{(1-\chi_s)/2}\eps(I_s) \theta_{a_s},
\eeq
$s\in S$, $\eps[g]\equiv\sign\det(g_{MN})$. More explicitly
(\ref{1.22}) reads: $\eps_s=\eps(I_s) \theta_{a_s}$ for
$v_s = e$ and $\eps_s=-\eps[g] \eps(I_s) \theta_{a_s}$, for
$v_s = m$.

Equations  (\ref{3.2a})  are equivalent to Toda-type
equations. First boundary condition guarantees the existence
of a regular horizon at $R^{\bar{d}} =   2 \mu$. Second
condition (\ref{3.2b}) ensures an asymptotical
(for $R \to +\infty$) flatness of the $(2+d_1)$-dimensional
section of the metric.

Due to (\ref{2.32}) and  (\ref{2.33}), the dimension of
$p$-brane worldsheet $d(I_s)$ is defined by
\ber{1.16a}
d(I_s)=  n_{a_s}-1, \quad d(I_s)=   D- n_{a_s} -1,
\eer
for $s \in S_e, S_m$ respectively.
For a $p$-brane: $p =   p_s =   d(I_s)-1$.

The solutions are valid if the following  restriction on the sets
$\Omega_{a,v}$ is imposed.
(This restriction guarantees the block-diagonal structure
of the stress-energy tensor, like for the metric, and the existence of
$\sigma$-model representation \cite{IMC}, see also \cite{AR}).
We denote $w_1\equiv\{i|i\in \{2,\dots,n\},\quad d_i=1\}$, and
$n_1=|w_1|$ (i.e. $n_1$ is the number of 1-dimensional spaces among
$M_i$, $i=2,\dots,n$).  It follows from  (\ref{1.7a})
that $2 \in w_1$.

{\bf Restriction.} {\em Let 1a) $n_1\le1$ or 1b) $n_1\ge2$ and for
any $a\in\tri$, $v\in\{e,m\}$, $i,j\in w_1$, $i \neq j$, there are no
$I,J\in\Omega_{a,v}$ such that $i \in I$, $j\in J$ and $I\setminus\{i\}=
J\setminus\{j\}$.}

These restriction is  satisfied in the non-composite case
\cite{IM0,IM}:  $|\Omega_{a,e}| + |\Omega_{a,m}| = 1$,
(i.e when there are no two  $p$-branes with the same color index $a$,
$a\in\tri$.) The restriction forbids certain intersections of two
$p$-branes with the same color index for  $n_1 \geq 2$.

The Hawking temperature corresponding to
the solution is (see also \cite{Oh,BIM} for orthogonal case)
found to be
 \beq{2.36}
T_H=   \frac{\bar{d}}{4 \pi (2 \mu)^{1/\bar{d}}}
\prod_{s \in S} H_{s0}^{- h_s},
\eeq
where $H_{s0}$ are defined in (\ref{3.2a})

This solution describes a set of charged (by forms) overlapping
$p$-branes ($p_s=d(I_s)-1$, $s \in S$) ``living'' on submanifolds
of $M_2 \times \dots \times M_n$.

\section{Examples.}

\subsection{Orthogonal and block-orthogonal solutions}

There exist solutions to eqs. (\ref{3.1})-(\ref{3.2b})
of polynomial type. The simplest example occurs in orthogonal
case \cite{CT,AIV,Oh,IMJ,BIM}, when
\beq{3.4}
B_{s s'} = 0,
\eeq
for  $s \neq s'$, $s, s' \in S$. In this case
$(A_{s s'}) = {\rm diag}(2,\ldots,2)$ is a Cartan matrix
for semisimple Lie algebra $A_1 \oplus  \ldots  \oplus  A_1$
and
\beq{3.5}
H_{s}(z) = 1 + P_s z,
\eeq
with $P_s \neq 0$, satisfying
\beq{3.5a}
P_s(P_s + 1) = - B_s,
\eeq
$s \in S$. Relation (\ref{3.2a}) implies $P_s > -1$.
For $B_s < 0$  parameters $P_s >0$ are uniquely defined.

In \cite{Br,IMJ2,CIM} this solution
was generalized to a block orthogonal
case:
\ber{3.6}
S=S_1 \cup\dots\cup S_k, \qquad  S_i \cap S_j = \emptyset, \quad i \neq j,
\eer
$S_i \ne \emptyset$, i.e. the set $S$ is a union of $k$ non-intersecting
(non-empty) subsets $S_1,\dots,S_k$,
and relation (\ref{3.4}) should be satisfied
for all $s\in S_i$, $s'\in S_j$, $i\ne j$; $i,j=1,\dots,k$.
In this case (\ref{3.5}) is modified as follows
\beq{3.8}
H_{s}(z) = (1 + P_s z)^{b_0^s},
\eeq
where
\beq{3.11}
b_0^s = 2 \sum_{s' \in S} A^{s s'},
\eeq
$(A^{s s'}) = (A_{s s'})^{-1}$ and parameters $P_s$ are coinciding inside
blocks, i.e. $P_s = P_{s'}$ for $s, s' \in S_i$, $i =1,\dots,k$.
 Parameters $P_s \neq 0 $ satisfy the relations

  $$P_s(P_s + 2\mu) = - \bar B_s/b_0^s,$$

$b_0^s \neq 0$, and parameters $\bar B_s/b_0^s$  are also
coinciding inside blocks, i.e. $\bar B_s/b_0^s = \bar
B_{s'}/b_0^{s'}$ for $s, s' \in S_i$, $i =1,\dots,k$.

Let $(A_{s s'})$ be  a Cartan matrix  for a  finite-dimensional
semisimple Lie  algebra $\cal G$. In this case all powers in
(\ref{3.11})  are  natural numbers  \cite{GrI} (coinciding with
the components  of twice the dual Weyl vector in the basis of
simple coroots, see \cite{FS}) and  hence, all functions $H_s$ are
polynomials, $s \in S$.

{\bf Conjecture.} {\em Let $(A_{s s'})$ be  a Cartan matrix
for a  semisimple finite-dimensional Lie algebra $\cal G$.
Then  the solution to eqs. (\ref{3.1})-(\ref{3.2b})
(if exists) is a polynomial
\beq{3.12}
H_{s}(z) = 1 + \sum_{k = 1}^{n_s} P_s^{(k)} z^k,
\eeq
where $P_s^{(k)}$ are constants,
$k = 1,\ldots, n_s$, integers $n_s = b_0^s$ are
defined in (\ref{3.11}) and $P_s^{(n_s)} \neq 0$,  $s \in S$}.

For certain series of simple finite-dimensional Lie algebras this
conjecture will be proved in a separate publication.  In the extremal case
$\mu = + 0$ an a analogue of this conjecture was suggested previously in
\cite{LMMP}.

\subsection{Solutions for $A_2$ algebra}

Here we consider some examples of solutions related
to the Lie algebra $A_2 = sl(3)$. In this case the Cartan
matrix reads
\beq{B.1a}
\left(A_{ss'}\right)=
\left( \begin{array}{*{6}{c}}
2&-1\\
-1&2\\
\end{array}
\right)\quad
\eeq

According to the
results of previous section we  seek the solutions
to eqs. (\ref{3.1})-(\ref{3.2b}) in the following
form (see  (\ref{3.12}); here $n_1 = n_2 =2$):
\beq{4.1}
H_{s} = 1 + P_s z + P_s^{(2)} z^{2},
\eeq
where $P_s= P_s^{(1)}$ and $P_s^{(2)} \neq 0$ are constants,
$s = 1,2$.

The substitution of  (\ref{4.1}) into equations  (\ref{3.1})
and decomposition in powers of $z$ lead us to the relations
\bear{4.2}
 - P_s (P_s + 1 )  + 2 P_s^{(2)} =  B_s, \\  \label{4.3}
 - 2 P_s^{(2)} (P_s + 2 ) =  P_{s+1} B_s, \\ \label{4.4}
 - 2 P_s^{(2)} ( \frac{1}{2} P_s + P_{s}^{(2)}) =  P_{s+1}^{(2)}  B_s,
\ear
corresponding to powers $z^0, z^1, z^2$ respectively, $s = 1,2$.
Here we denote  $s+ 1 = 2, 1$ for $s = 1,2$ respectively.
For $P_1 +P_2 + 2 \neq 0$ the solutions to eqs.
(\ref{4.2})-(\ref{4.4}) read
\bear{4.5}
 P_s^{(2)} = \frac{ P_s P_{s +1} (P_s + 1 )}{2 (P_1 +P_2 + 2)},
 \\ \label{4.6}
B_s = - \frac{ P_s (P_s + 1 )(P_s + 2 )}{P_1 +P_2 + 2},
\ear
$s = 1,2$. For $P_1 +P_2 + 2 = 0$ there exist also a
special solution with
\bear{4.6a}
P_1= P_2 = -1 , \qquad  2 P_{s}^{(2)} =  B_s >0,
\qquad  B_1 +  B_2 = 1.
\ear

Thus, in the $A_2$-case the solution is described by relations
(\ref{2.30})-(\ref{2.33}) with $S = \{s_1,s_2\}$,
intersection rules following from (\ref{1.17}), (\ref{1.18})
and (\ref{B.1a})
\bear{1.40a}
d(I_{s_1} \cap I_{s_2})= \frac{d(I_{s_1})d(I_{s_2})}{D-2}-
\chi_{s_1} \chi_{s_2} \lambda_{a_{s_1}}\cdot\lambda_{a_{s_2}}
- \frac12 K,
\\ \label{1.40b}
d(I_{s_i}) - \frac{(d(I_{s_i}))^2}{D-2}+
\lambda_{a_{s_i}}\cdot\lambda_{a_{s_i}} = K,
\ear
where
$K = K_{s_i} \neq 0$,  and functions $H_{s_i} = H_i$
are defined by relations
(\ref{4.1}) and (\ref{4.5})-(\ref{4.6a}) with $z = 2\mu R^{-\bar d}$,
$i =1,2$. Here $\lambda\cdot\lambda^{'}=
\lambda_{\alpha}\lambda_{\beta }^{'}h^{\alpha\beta}$.

\subsection{$A_2$-dyon in $D = 11$ supergravity }

Consider the  ``truncated''  bosonic sector of
$D=   11$ supergravity (``truncated''  means without
Chern-Simons term).  The action  (\ref{1.1}) in this
case reads  \cite{CJS}
\ber{4.7}
S_{tr} =   \int_{M} d^{11}z \sqrt{|g|}
\left\{ {R}[g] - \frac{1}{4!}  F^2 \right\}.
\eer
where ${\rm rank} F =   4$. In this particular case,
we consider a dyonic black-hole solutions
with  electric $2$-brane and magnetic  $5$-brane
defined on the manifold
\beq{4.8}
M =    (2\mu, +\infty )  \times
(M_1 = S^{2})  \times (M_2 = \R) \times M_{3} \times M_{4},
\eeq
where ${\dim } M_3 =  2$ and ${\dim } M_4 =  5$.

The solution reads,
\bear{4.9}
g=  H_1^{1/3} H_2^{2/3} \left\{
f^{-1}dR \otimes dR + R^2 d \Omega^2_2
 -  H_1^{-1} H_2^{-1} f dt\otimes dt
+ H_1^{-1} g^3 + H_2^{-1} g^4 \right\}, \\
\label{4.10}
F =  - \frac{Q_1}{R^2} H_1^{-2} H_2  dR \wedge dt \wedge \tau_3+
Q_2 \tau_1 \wedge \tau_3,
\ear
where $f = 1 - 2\mu/R$, metrics $g^2$ and  $g^3$ are
Ricci-flat metrics of Euclidean signature,
and  functions $H_{s}$  are defined by relations (\ref{4.1}),
(\ref{4.5}) and (\ref{4.6}) with $z = 2\mu R^{- 1}$,
$B_s = - 2 Q_s^2/(2\mu)^2$, $s =1, 2$;
where $\tau_1$ is volume form on $S^2$.

The  solution describes $A_2$-dyon consisting
of electric  $2$-brane with worldsheet isomorphic
to $(M_2 = \R) \times M_{3}$ and magnetic  $5$-brane
with worldsheet isomorphic to $(M_2 = \R) \times M_{4}$.
The ``branes'' are intersecting on the time manifold $M_2 = \R$.
Here  $K_s = (U^s,U^s)=2$, $\eps_s = -1$ for all $s \in S$.
The $A_2$ intersection rule reads (see (\ref{1.40a}))
\beq{4.12}
2 \cap 5= 1
\eeq
Here and in what follows $(p_1 \cap p_2= d) \Leftrightarrow
(d(I)=p_1 + 1, d(J)= p_2 + 1, d(I\cap J) = d)$.

The solution (\ref{4.9}), (\ref{4.10})
satisfies not only equations of
motion for the truncated model,
but also  the equations of motion
for  $D =11$ supergravity with the bosonic sector action
\ber{4.13}
S =  S_{tr} +  c \int_{M} A \wedge F \wedge F
\eer
($c = {\rm const}$,  $F = d A$),
since the only modification
related to ``Maxwells'' equations
\ber{4.14}
d*F = {\rm const} \ F \wedge F,
\eer
is trivial due to $F \wedge F = 0$ (since $\tau_i \wedge \tau_i =0$).

This solution in a special case $H_1 = H_2 = H^2$
($P_1 = P_2$, $Q_1^2 =  Q_2^2$) was considered in \cite{CIM}.
The 4-dimensional section of the metric (\ref{4.9})
in this special case coincides
with the Reissner-Nordstr\"om  metric.
For the extremal case, $\mu \to + 0$,
and multi-black-hole generalization
see also  \cite{IMBl}.

\subsection{$A_2$-dyon in Kaluza-Klein model}

Let us  consider $4$-dimensional model
\beq{4.15}
S= \int_{M} d^4z \sqrt{|g|}\biggl\{R[g]- g^{\mu \nu}
\p_\mu \varphi \p_\nu \varphi
-\frac{1}{2!} \exp[2\lambda \varphi]F^2\biggr\}
\eeq
with scalar field  $\varphi$, two-form $F = d A$ and
$\lambda = - \sqrt{3/2}$.
This model originates after Kaluza-Klein (KK) reduction
of $5$-dimensional gravity. The 5-dimensional metric
in this case reads
\beq{4.16a}
g^{(5)} = \phi g_{\mu \nu} dx^{\mu} \otimes dx^{\nu}
          + \phi^{-2} (dy + {\cal A}) \otimes (dy + {\cal A}),
\eeq
where ${\cal A} = \sqrt{2} A =  \sqrt{2} A_{\mu} dx^{\mu}$ and
$\phi = \exp(2 \varphi/\sqrt{6})$.

We consider a dyonic black-hole solution
carrying  electric  charge $Q_1$
and magnetic  charge $Q_2$, defined on the manifold
$M =    (2\mu, +\infty )  \times (M_1 =S^{2})  \times (M_2 = \R)$.
This solution reads
\bear{4.18}
g= \left( H_1 H_2 \right)^{1/2}
\biggl\{ \frac{dR \otimes dR}{1 - 2\mu / R} +   R^2  d \Omega^2_{2}
 -  H_1^{-1} H_2^{-1} \left(1 - \frac{2\mu}{R } \right) dt\otimes dt
\biggr\}, \\ \label{4.19}
\exp(\varphi) = H_1^{\lambda/2} H_2^{- \lambda/2},
\\ \label{4.20}
F = dA = -  \frac{Q_1}{R^2} H_1^{-2} H_2  dR \wedge dt + Q_2 \tau_1,
\ear
where  functions $H_{s}$  are defined by relations (\ref{4.1}),
(\ref{4.5}) and (\ref{4.6}) with $z = R^{- 1}$,
$B_s = - 2 Q_s^2/(2\mu)^2$, $s =1, 2$;
where $\tau_1$ is volume form on $S^2$.

For 5-metric we obtain from (\ref{4.16a})-(\ref{4.19})
\bear{4.16c}
g^{(5)} = H_2 \biggl\{ \frac{dR \otimes dR}{1 - 2\mu / R}
+   R^2  d \Omega^2_{2}   -  H_1^{-1} H_2^{-1}
\left(1 - \frac{2\mu}{R } \right) dt\otimes dt \biggr\}
\\ \nn
 + H_1 H_2^{-1}  (dy + {\cal A}) \otimes (dy + {\cal A}),
\ear
$d {\cal A} = \sqrt{2} F$.

For $Q_2 \to 0$ we get the black hole version of
Dobiash-Maison solution from \cite{DoMa} and
for $Q_1 \to 0$ we are led to the black hole version of
Gross-Perry-Sorkin monopole solution from \cite{GrP,Sor},
(see \cite{CGMS}).  The solution coincides  with
Gibbons-Wiltshire dyon solution  \cite{GW}. Our notations are related to
those from ref. \cite{GW}, as following :  $H_1 R^2 = B$, $H_2 R^2 = A$,
$R^2 - 2 \mu R = \Delta$, $Q_1 = \sqrt{2} q$, $Q_2 = - \sqrt{2} p$, $R
-\mu = r- m$, $\mu^2 = m^2 + d^2 - p^2 -q^2$, $(P_2 -P_1)/2(P_2 +1) = d/(d
-\sqrt{3} m)$).  (For general spherically symmetric configurations see
also ref. \cite{Lee}.)

We note that, quite recently, in \cite{CGS} the KK dyon solution
\cite{GW} was used for constructing the dyon solution in $D=11$
supergravity (\ref{4.9})-(\ref{4.10}) for flat $g^3$ and $g^4$ and its
rotating version.  This dyon solution differs from
$M2 \subset M5$ dyon from \cite{ILPT,GLPT,Co}.

\section{Conclusions}

Thus here we presented a family
of black hole solutions  with intersecting $p$-branes
with next to arbitrary intersection rules (see
(\ref{1.17a}), (\ref{1.18b}) and Restriction).
The metric of solutions  contains  $n -1$ Ricci-flat ``internal''
space metrics. The solutions are defined up to  a set of functions $H_s$
obeying a set of  equations (equivalent to Toda-type
equations) with certain boundary conditions imposed.
Using  a conjecture on polynomial structure of  $H_s$
for intersections related to semisimple Lie algebras,
we obtained explicit relations for the solutions in the $A_2$-case
and considered two examples of $A_2$-dyon solutions: one in $D =
11$ supergravity (with M2 and M5 branes intersecting at a point)
and another in $5$-dimensional Kaluza-Klein theory.

\begin{center}
{\bf Acknowledgments}
\end{center}

This work was supported in part by the Russian Ministry for
Science and Technology, Russian Foundation for Basic Research,
and project SEE.

\end{document}